\documentclass{iopart}

\usepackage{iopams}
\usepackage[dvips]{graphicx}

\begin{document}

\title[Statistical $J/\psi$ production and open charm enhancement...]{Statistical 
$J/\psi$ production and open charm enhancement in Pb+Pb collisions at CERN SPS}

\author{A.P. Kostyuk}

\address{Institut f\"ur Theoretische Physik, Universit\"at  Frankfurt,
Germany\\
and\\
Bogolyubov Institute for Theoretical Physics,
Kyiv, Ukraine}

\ead{kostyuk@th.physik.uni-frankfurt.de}

\begin{abstract}
Production of open and hidden charm hadrons in heavy ion collisions is 
considered within the statistical coalescence model. Charmed 
quarks and antiquarks are assumed to be created at the initial stage 
of the reaction and their number is conserved during the evolution of 
the system. They are distributed among open and hidden charm hadrons 
at the hadronization stage in accordance with  
laws of statistical mechanics. The model is in excellent agreement with 
the experimental data on $J/\psi$ production in lead-lead collisions at 
CERN SPS and predicts strong 
enhancement of the open charm multiplicity over the standard 
extrapolation from nucleon-nucleon to nucleus-nucleus collisions.
A possible mechanism of the charm enhancement is proposed. 
\end{abstract}


$J/\psi$ meson plays a special role in heavy ion physics.
The interest to it was mainly motivated by the suggestion 
of Matsui and Satz \cite{MS}
to use charmonia as a probe of the state of matter created 
in the early stage of the collision. 

The standard picture assumes that
charmonia are created exclusively at the initial stage of the 
reaction in primary nucleon-nucleon collisions. During the subsequent
evolution of the system, the number of hidden charm mesons is reduced
because of (a) absorption of pre-resonance charmonium states in the nuclei 
(normal nuclear suppression),
(b) interactions of charmonia with secondary hadrons (comovers), 
(c) dissociation of $c\bar{c}$ bound states in deconfined medium.
It was found that the $J/\psi$ suppression with respect to
Drell-Yan muon pairs measured in proton-nucleus and 
nucleus-nucleus collisions with light projectiles 
can be explained by normal nuclear suppression alone \cite{NA38}.
In contrast, the NA50 experiment with a heavy projectile and target
(lead-lead) revealed essentially stronger $J/\psi$ suppression for central
collisions \cite{anomalous}. 
This {\it anomalous} $J/\psi$ suppression was attributed to
formation of quark-gluon plasma \cite{evidence}.   

\begin{figure}[t]
\begin{center}
\includegraphics[height=8cm]{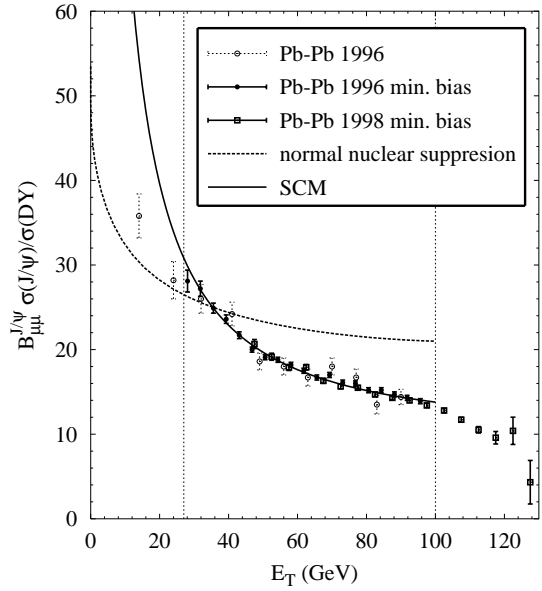} 
\caption{The dependence of the $J/\psi$ to Drell-Yan 
ratio on the transversal energy. The normal nuclear suppression curve 
is obtained at $\sigma_{abs} = 6.4$~mb, where $\sigma_{abs}$ is the 
absorption cross section of preresonant charmonia by nuclear 
nucleons. Two vertical lines show the applicability domain of the statistical
coalescence model (SCM), see \cite{Ko:01} for details. 
\label{Jpsi_PbPb} }
\end{center}
\end{figure}

Much less attention was paid to the fact that the $J/\psi$ to Drell-Yan
ratio {\it does not} follow the normal nuclear suppression pattern  for 
{\it peripheral} collisions: the experimental points 
lie {\it above} the normal nuclear suppression curve (see Fig. \ref{Jpsi_PbPb}). 
This behavior cannot be explained within the standard scenario:
presence of either of two additional mechanisms (b) or (c) can only
destroy the quarkonia that survived absorption by the nuclear nucleons.

A completely different picture of charmonium production was proposed
by  Ga\'zdzicki and Gorenstein \cite{GG}: hidden charm mesons 
are supposed to 
be created at the hadronization stage. Similar to all other hadrons,
their abundancies can be described within the thermal model \cite{thermal}.
However, production of heavy quarks in soft processes is expected
to be negligible. Most likely, 
they are produced at the hard stage. Therefore, their number can, generally 
speaking, deviate from the thermal equilibrium value \cite{Br1}.
Because of smallness of this number, canonical treatment of the system 
is important \cite{Go:00}.   

In this talk I present the results \cite{Ko:01} obtained within the 
statistical coalescence model \cite{Go:00}:
\begin{itemize}
\item 
$c$ and $\bar{c}$ are created at the 
initial stage of the reaction in primary hard parton collisions;
\item
their number remains approximately unchanged during
the subsequent evolution;
\item
they are distributed over open charm hadrons and charmonia
at the hadronization stage in accordance with
laws of statistical mechanics. 
\end{itemize}


Within its applicability domain,
the model demonstrates excellent agreement  ($\chi^2/\mbox{dof} = 1.16$)
with the NA50 data on $J/\psi$ production (see figure \ref{Jpsi_PbPb}).

Extrapolating the fit to peripheral collisions reveals an interesting
feature: our model predicts sharp increase 
of the ratio with decreasing the number of participants, leading to $J/\psi$
{\it enhancement} over the nucleon-nucleon collision value. Such 
behavior is easy to understand: the smaller is the volume of the 
system the larger is the 
probability that $c$ and $\bar{c}$ meet each other at the hadronization
stage and form a hidden charm meson. 
As is seen from the figure, this enhancement is not obviously
supported by the data: our theoretical curve 
lies above the experimental points in the low $E_T$ region.
On the other hand, the normal nuclear suppression model also fails
to explain the leftmost point from the 1996 standard analysis set
and two leftmost points from the 1996 minimum bias set.
The theoretical calculations underestimate the experimental
values.  It is natural to assume that an intermediate situation
takes place. Some fraction of peripheral Pb+Pb collisions result in
formation of deconfined medium. In these collisions, charmonia are 
formed at the hadronization stage, and their multiplicities are given 
by the statistical coalescence model. 
The rest collisions (we shall call them 'normal collisions') 
do not lead to color deconfinement, therefore charmonia are formed
exclusively at the initial stage and then suffer 
normal nuclear suppression.
The experiment measures the average value, which lies between these
two curves. The fraction of 'normal' events decreases with growing centrality. 
Their influence on $J/\psi$ production becomes negligible
at $N_p \gtrsim 100$. 

The model has two free parameters: $\sigma^{NN}_{c\bar{c}}$, 
the effective cross section of 
charm production by a nucleon pair, and
$\eta$,  the fraction of $J/\psi$ satisfying the kinematical conditions 
of the NA50 spectrometer.

Our analysis predicts enhancement of the total
charm by a factor of $4.5$--$7.5$ 
($\sigma^{NN}_{c\bar{c}}=34 \pm 9$ $\mu$b,
comparing with $\sigma^{NN}_{c\bar{c}} \approx 5.5$ $\mu$b
in nucleon-nucleon collisions.
On the other hand, the small value of $\eta \approx 0.14$ 
(comparing with $\eta \approx 0.24$ in nucleon-nucleon collisions) 
suggest essential
broadening of the $J/\psi$ rapidity distribution.
Assuming approximately the same broadening for the open charm, we obtain
the open charm enhancement by a factor of about $2.5$--$4.5$ within the 
rapidity window of the NA50 spectrometer.
This is consistent with the indirect experimental result \cite{NA50open}.

A direct measurement of open charm in heavy ion collisions is 
planed at CERN SPS.
If the future measurement confirms the enhancement, 
what kind of mechanism can stay behind this 
phenomenon? 

We have demonstrated \cite{hf_enh} that the heavy flavor enhancement may be a 
signature of color deconfinement.
Deconfined medium (quark-gluon plasma or its precursor) can
influence hadronization of heavy  
quarks and antiquarks.
This leads to enhanced production of hadrons with open charm 
and bottom with respect to the direct extrapolation of proton-proton data to
heavy nuclei collisions.

To get an intuitive picture of possible medium effects let us 
start from open charm production in electron-positron annihilation.
A charm quark-antiquark pair is created in a hard perturbative 
process. Then it hadronizes into observed particles. The dynamics of 
the hadronization   can be qualitatively understood in the framework 
of the string picture. When the distance between $c$ and 
$\overline{c}$ reaches the range of confinement forces, 
a string connecting these colored objects is formed.
If the invariant mass of the quark-antiquark pair
$M_{c\overline{c}}$
lies well above the open charm meson threshold $2m_D$,  
$c$ and $\overline{c}$ break 
the string into two (or more) peaces, so that the final state 
contains an open charm hadron pair (and possibly a number of light hadrons).
However, when the center-of-mass energy of the initial electron-positron
pair exceeds the charm quark  threshold ($\sqrt{s} > 2 m_c$),
but lies below the $D$-meson threshold
($\sqrt{s} < 2 m_D$), a $c\overline{c}$ pair can be created, but it cannot break
the string. An open charm hadron pair  {\it cannot} be formed. 
Eventually the $c$ and $\overline{c}$ have to annihilate into lighter
hadrons (or form charmonium states, 
provided that the energy is sufficient). 

Let us imagine now $e^+e^-$ annihilation inside
deconfined medium.
Due to the Debye screening, no string is formed between colored 
objects. If $c$ and $\overline{c}$ are created,
they {\it can} fly apart within the medium as 
if they were free particles. It does not matter whether their
initial invariant mass $M_{c\overline{c}}$ exceeds the 
$D$-meson threshold or not. 
The created $c\overline{c}$ pair {\it is able} to form  
an open charm hadron pair at the hadronization.
This means that {\it 
$e^+e^-$ annihilation inside deconfined medium would
produce open charm hadrons even if the collision energy is not sufficient
for producing these hadrons in vacuum.  
}

In proton-proton or nucleus-nucleus collisions charm quark-antiquark  
pairs are produced due to hard {\it parton} interactions.
Calculations in the leading order of perturbative 
quantum chromodynamics show  that 
a great fraction of $c\overline{c}$ pairs is created 
with invariant masses $M_{c\overline{c}}$ below the 
corresponding meson threshold $2m_D$.
Despite of essential differences between $e^+e^-$ annihilation and
nucleon-nucleon (or nucleus-nucleus) collisions 
(see \cite{hf_enh} for details), it is natural to expect that
deconfined medium essentially increase the probability of hadronization 
of $c\overline{c}$ pairs with low invariant mass.
{\it  This should lead to enhanced production of the total charm 
in nucleus-nucleus collisions in comparison to the standard result obtained 
within the direct extrapolation of nucleon-nucleon data.}

We have estimated the upper bound of this enhancement. At SPS energies,
one can expect  maximal  enhancement by a factor of about 6,
which is consistent
with our above fit for the $J/\psi$ to Drell-Yan ratio.

I conclude that the NA50 data for
not very peripheral lead-lead collisions  are consistent 
with the following scenario:

Formation of deconfined medium prevents annihilation of charm-anticharm
quark pairs with low invariant mass. This reveals itself in enhanced 
production of open charm hadrons. 
Charmonia as well as other hadrons are 
formed at the hadronization stage. 
The distribution
of charm quarks and antiquarks over open and hidden 
charm hadrons follows laws of statistical mechanics. 

\ack
I am indebted to M.~Gorenstein, 
W.~Greiner,  L.~McLerran and H.~St\"ocker for fruitful
collaboration.  
I wish to thank F.~Becattini, P.~Bordalo, D.~Blaschke, L.~Bravina, 
K.~Bugaev, M.~Ga\'zdzicki, L.~Gerland, J.~Cleymans, B.~K\"ampfer, 
P.~Minkowski, J.~Rafelski, D.~Rischke, I.~Shovkovy and E.~Zabrodin
for useful comments and interesting discussions.
The research described in this publication was made possible in part by
Award \# UP1-2119 of the U.S. Civilian Research and Development
Foundation for the Independent States of the Former Soviet Union
(CRDF).


\section*{References}

\begin{thebibliography}{99}

\bibitem{MS}
Matsui T and Satz H 1986
{\it Phys.\ Lett.}\ B {\bf 178}  416\\
Satz H 2000
{\it Rept.\ Prog.\ Phys.}\  {\bf 63} 1511

\bibitem{NA38}
Abreu M C {\it et al.} 1999
{\it Phys.\ Lett.}\ B {\bf 466} 408

\bibitem{anomalous}
Abreu M C  {\it et al.}  (NA50 Collaboration) 1997
{\it Phys.\ Lett.}\ B {\bf 410}  337\\
Abreu M C  {\it et al.}  (NA50 Collaboration) 1999
{\it Phys.\ Lett.}\ B {\bf 450} 456

\bibitem{evidence}
Abreu M C  {\it et al.}  (NA50 Collaboration) 2000
{\it Phys.\ Lett.}\ B {\bf 477} 28

\bibitem{GG}
Ga\'zdzicki M and Gorenstein  M I 1999
{\it Phys.\ Rev.\ Lett.}\  {\bf 83}  4009


\bibitem{thermal}
Braun-Munzinger~P, Heppe~I and Stachel~J 1999
{\it Phys.\ Lett.}\ B {\bf 465}  15\\
Becattini~F, Cleymans~J, Keranen~A, Suhonen~E and Redlich~K 2001
{\it Phys.\ Rev.}\ C {\bf 64} 024901\\
Yen G D and Gorenstein M I 1999 {\it Phys. Rev}. C {\bf 59} 2788



\bibitem{Br1}
Braun-Munzinger P and Stachel J 2000
{\it Phys.\ Lett.}\ B {\bf 490} 196 

\bibitem{Go:00}
Gorenstein~M~I, Kostyuk~A~P, Stocker~H and Greiner~W 2001
{\it Phys.\ Lett.}\ B {\bf 509} 277\\
Gorenstein~M~I, Kostyuk~A~P, Stocker~H and Greiner~W 2001
{\it J.\ Phys.}\ G {\bf 27} L47

\bibitem{Ko:01}
Kostyuk~A~P, Gorenstein~M~I, St\"ocker~H and Greiner~W 2001
{\it Preprint} hep-ph/0110269

\bibitem{NA50open}
M.~C.~Abreu {\it et al.}  (NA38 and NA50 Collaborations) 2000
{\it Eur.\ Phys.\ J.}\ C {\bf 14} 443

\bibitem{hf_enh}
Kostyuk~A~P, Gorenstein~M~I and Greiner W 2001
{\it Phys.\ Lett.}\ B {\bf 519} 207

\end{thebibliography}
\end{document}